\def\version{October 14, 2006}

\newif\ifpdf
\ifx\pdfoutput\undefined
\pdffalse 
\else
\pdfoutput=1 
\pdftrue \fi

\newif\iffinal
\finaltrue 

\documentclass[reqno,11pt]{amsart}
\iffinal\else\usepackage[notref,notcite]{showkeys}\fi
\iffinal\else\IfFileExists{pdfsync.sty}{\usepackage{pdfsync}}{}\fi
\usepackage{amsmath}
\usepackage{amsfonts}
\usepackage{amssymb}
\usepackage{verbatim}
\usepackage{epsfig}

\IfFileExists{epsf.def}{\input epsf.def}{\usepackage{epsf}}

\IfFileExists{myowntimes.sty}{\usepackage[itboldsymbol]{myowntimes}\usepackage{mathrsfs}}
{\usepackage{times}\IfFileExists{mathrsfs.sty}
{\usepackage{mathrsfs}}{\newcommand{\mathscr}{\mathcal}}}
\DeclareFontFamily{OT1}{eusb}{} \DeclareFontShape{OT1}{eusb}{m}{n}
{<5> <6> <7> <8> <9> <10> <11> <12> <14.4> eusb10}{}
\DeclareMathAlphabet{\eusb}{OT1}{eusb}{m}{n}

\DeclareFontFamily{OT1}{eusm}{} \DeclareFontShape{OT1}{eusm}{m}{n}
{<5> <6> <7> <8> <9> <10> <11> <12> <14.4> eusm10}{}
\DeclareMathAlphabet{\eusm}{OT1}{eusm}{m}{n}

\DeclareFontFamily{OT1}{eufm}{} \DeclareFontShape{OT1}{eufm}{m}{n}
{<5> <6> <7> <8> <9> <10> <11> <12> <14.4> eufm10}{}
\DeclareMathAlphabet{\mathfrak}{OT1}{eufm}{m}{n}

\DeclareFontFamily{OT1}{fraktura}{}
\DeclareFontShape{OT1}{fraktura}{m}{n} {<5> <6> <7> <8> <9> <10>
<11> <12> <13> <14.4> [1.1] eufm10}{}
\DeclareMathAlphabet{\fraktura}{OT1}{fraktura}{m}{n}

\DeclareFontFamily{OT1}{cmfi}{} \DeclareFontShape{OT1}{cmfi}{m}{n}
{<5> <6> <7> <8> <9> <10> <11> <12> <13> <14.4> [0.9] cmfi10}{}
\DeclareMathAlphabet{\cmfi}{OT1}{cmfi}{b}{n}

\DeclareFontFamily{OT1}{cmss}{} \DeclareFontShape{OT1}{cmss}{m}{n}
{<5> <6> <7> <8> <9> <10> <11> <12> <13> <14.4> cmss10}{}
\DeclareMathAlphabet{\cmss}{OT1}{cmss}{m}{n}

\newtheoremstyle{thm}{1.5ex}{1.5ex}{\itshape\rmfamily}{} {\bfseries\rmfamily}{}{2ex}{}

\newtheoremstyle{def}{1.5ex}{1.5ex}{\slshape\rmfamily}{} {\bfseries\rmfamily}{}{2ex}{}

\newtheoremstyle{rem}{1.3ex}{1.3ex}{\rmfamily}{} {\itshape}
{} {1.5ex}{}

\theoremstyle{thm}
\newtheorem{theorem}{Theorem}[section]
\newtheorem{lemma}[theorem]{Lemma}
\newtheorem{proposition}[theorem]{Proposition}

\newtheorem*{Main Theorem}{Main Theorem.}

\newtheorem*{special theorem}{Lindeberg-Feller Theorem for Martingales}

\theoremstyle{def}

\theoremstyle{rem}
\newtheorem{remark}[theorem]{{\itshape Remark}}

\numberwithin{equation}{section}

\renewcommand{\section}{\secdef\sct\sect}
\newcommand{\sct}[2][default]{%
\refstepcounter{section}
\addcontentsline{toc}{section}{{\tocsection
{}{\thesection}{\!\!\!\!#1\dotfill}}{}}
\vspace{0.7cm}
\centerline{\scshape\thesection.\ #1} \nopagebreak \vspace{0.2cm}}
\newcommand{\sect}[1]{%
\vspace{0.4cm} \centerline{\large\scshape\rmfamily #1}
\vspace{0.2cm}}

\renewcommand{\subsection}{\secdef\subsct\sbsect}
\newcommand{\subsct}[2][default]{\refstepcounter{subsection}
\addcontentsline{toc}{subsection}
{{\tocsection{\!\!}{\hspace{1.2em}\thesubsection}{\!\!\!\!#1\dotfill}}{}}
\nopagebreak\vspace{0.45\baselineskip} {\flushleft\bf
\thesubsection~\bf #1.~}
\\*[3mm]\noindent
\nopagebreak}
\newcommand{\sbsect}[1]{\vspace{0.1cm}\noindent
\textbf{#1.~}\vspace{0.1cm}}

\renewcommand{\subsubsection}{%
\secdef \subsubsect\sbsbsect}
\newcommand{\subsubsect}[2][default]{%
\refstepcounter{subsubsection}
\addcontentsline{toc}{subsubsection}{{\tocsection{\!\!}
{\hspace{3.05em}\thesubsubsection}{\!\!\!\!#1\dotfill}}{}}
\nopagebreak
\vspace{0.15\baselineskip} \nopagebreak {\flushleft\rmfamily
\itshape\thesubsubsection
\ \rmfamily #1\/.}\ }
\newcommand{\sbsbsect}[1]{\vspace{0.1cm}\noindent
\rmfamily \itshape
\arabic{section}.\arabic{subsection}.\arabic{subsubsection} \
\sffamily #1\/.\ }

\iffinal
\newcommand{\printversion}{}
\else
\newcommand{\printversion}{,
\version}
\fi

\renewcommand{\caption}[1]{%
\vglue0.5cm
\refstepcounter{figure}
\begin{minipage}{0.9\textwidth}\small {\sc Figure~\thefigure. }#1\end{minipage}}

\newcommand{\textd}{\text{\rm d}\mkern0.5mu}

\newcommand{\1}{\text{\sf 1}}

\newcommand{\E}{\mathbb E}

\newcommand{\BbbP}{\mathbb P}

\def\myffrac#1#2 in #3{\raise 2.6pt\hbox{$#3 #1$}\mkern-1.5mu\raise 0.8pt\hbox{$#3/$}\mkern-1.1mu\lower 1.5pt\hbox{$#3 #2$}}
\newcommand{\ffrac}[2]{\mathchoice%
{\myffrac{#1}{#2} in \scriptstyle}
{\myffrac{#1}{#2} in \scriptstyle}
{\myffrac{#1}{#2} in \scriptscriptstyle}
{\myffrac{#1}{#2} in \scriptscriptstyle}
}

\newcommand{\bra}[1]{\langle#1|}
\newcommand{\ket}[1]{|#1\rangle}

\newcommand{\TR}{\text{\rm Tr}}

\title[Mean Field Quantum Spin Glasses\printversion]
{\fontsize{13}{16}\selectfont Thermodynamics and Universality for
Mean Field Quantum Spin Glasses}
\author[N.~Crawford\printversion]{Nicholas Crawford}

\begin{document}
\maketitle

\vspace{-5mm}
\centerline{\textit{Department of Mathematics, University of
California at Los Angeles}}

\begin{abstract}
We study aspects of the thermodynamics of quantum versions of spin
glasses. By means of the Lie-Trotter formula for exponential sums of
operators, we adapt methods used to analyze classical spin glass
models to answer analogous questions about quantum models.
\end{abstract}

\section{Introduction}

Classical spin glass models have seen a flurry of activity over the
last few years, see
\cite{A-S-S,C-H,Guerra,Guerra-Ton-1,Guerra-Ton-2,N-S-1,N-S-2,N-S-3,deSanc,
Talagrand-book,Talagrand-paper} to name a few; in particular
\cite{A-S-S, Guerra, Guerra-Ton-1, Talagrand-paper} all consider
aspects of the (generalized) Sherrington-Kirkpatrick model, a
mean-field model in which the interactions between spins are
mediated by an independent collection of Gaussian random variables.
In contrast, though physicists have considered both short and long
range quantum spin glass models for quite sometime, see \cite{B-H-G,
G-P-S, Rit, Sach-Ye}, there are few rigorous mathematical results;
we mention \cite{Con-Pul}, which provides a proof that the quenched
free energy of certain short ranged quantum spin glasses exists.

Here we extend classical spin glass results to quantum models in two
directions. First, using the ideas of \cite{Guerra-Ton-2}, we
demonstrate the existence of the quenched free energy of the
Sherrington-Kirkpatrick spin glass with a transverse external field.
Next, under conditions made precise below, we give a complementary
result which shows that a large class of quantum spin glasses,
including the transverse S-K model, satisfies universality. By this
we mean that as the size of the system goes to infinity, the
asymptotics of the free energy of the system do not depend on the
type of disorder used to define model.  This latter result is based
on the work of \cite{C-H}.

One may view this paper as an attempt to adapt the methods of
classical spin systems to the analysis of various quantum models.
Guerra's interpolation scheme \cite{Guerra, Guerra-Ton-1,
Guerra-Ton-2} and the Gaussian integration-by-parts formula are
ubiquitous tools in the classical setting.  A major theme of the
present work is that through the systematic use of the Lie-Trotter
product formula (i.e. that
\begin{equation}
e^{A+B}= \lim_{k \rightarrow \infty} \prod_{j=1}^k
e^{\frac{A}{k}}e^{\frac{B}{k}}
\end{equation}
for any pair of operators $A, B$) one may extend the interpolation
scheme and integration-by-parts formula to quantum systems in useful
ways.

For concreteness, the remainder of the introduction and the
following section use the language of the spin-$\ffrac 12$
representation of $\mathfrak{su}(2)$ to describe quantum spin
systems, though a number of our methods  apply in larger generality.
In this setting, one describes each particle using a two dimensional
Hilbert space $\mathbb C^2$ along with a representation of
$\mathfrak{su}(2)$ generated by the triplet of Pauli operators
$\vec{S}=(S^{(x)}, S^{(y)}, S^{(z)})$.

To represent an $N$-particle system, we  introduce the tensor
product $\mathbb V_N= \bigotimes_{j=1}^N \mathbb C^2$, one factor
for each particle, along with a sequence $(\vec{S}_j)_{j=1}^N$ of
$N$ copies of the Pauli vector $\vec {S}$, where $\vec{S}_j$ acts on
the $j$'th factor. The particles interact by means of the
Gibbs-Boltzmann operator $e^{-\beta \mathcal H_N}$ associated to the
Hamiltonian $\mathcal H_N$. Here, $\mathcal H_N$ is a self adjoint
operator acting on $\mathbb V_N$, typically a polynomial in the
$N$-tuple of spin operators $(\vec {S}_j)_{j=1}^N$. For example, the
Hamiltonian of the simplest non-trivial quantum spin system, the
transverse Ising model, is described by
\begin{equation}\label{SKT}
-\mathcal H_N = \frac{1}{N}\sum_{i,j=1}^N S^{(z)}_i S^{(z)}_j +
\lambda \sum_{j=1}^N S^{(x)}_j
\end{equation}
where $\lambda>0$.

Once we specify the Hamiltonian, statistical quantities of the
system may be defined.  The partition function and free energy of a
quantum spin system are defined via the trace of the Gibbs-Boltzmann
operator as
\begin{align*}
Z_N(\beta) = & \TR\left(e^{-\beta \mathcal H_N}\right)\\
f_N(\beta) = & \frac{-1}{N\beta} \log Z_N.
\end{align*}
Self adjoint operators on $\mathbb V_N$ replace functions as
observables of the system and the thermal average of an observable
$A$ is defined as
\begin{equation*}
\langle A \rangle = \frac{\TR\left(A e^{-\beta \mathcal
H_N}\right)}{\TR\left(e^{-\beta \mathcal H_N}\right)}.
\end{equation*}

With this formalism we present a few examples of spin glasses of
particular interest. Traditionally the modeling of any spin glass
necessitates the introduction of disordered interactions between
spin.  In general, and will be the case here, the interactions are
i.i.d. The basic example, the transverse S-K model, has a
Hamiltonian defined by
\begin{equation*}
-\mathcal H_N = \frac{1}{2 \sqrt N}\sum_{i,j=1}^N J_{i,j} S^{(z)}_i
S^{(z)}_j + \lambda \sum_{j=1}^N S^{(x)}_j.
\end{equation*}
A more complicated class of models, the quantum Heisenberg spin
glasses, are described by one of the Hamiltonians
\begin{align*}
-\mathcal {H}_N =& \frac{1}{2 \sqrt N}
\sum_{i,j=1}^NJ_{i,j}\left[S^{(z)}_i S^{(z)}_j+ \alpha
S^{(x)}_i S^{(x)}_j + \gamma S^{(y)}_i S^{(y)}_j \right] \\
-\mathcal {\tilde{H}}_N =& \frac{1}{2 \sqrt N} \sum_{\nu \in
\{x,y,z\}} \sum_{i,j=1}^N J^{(\nu)}_{i,j} S^{(\nu)}_i S^{(\nu)}_j,
\end{align*}
where $\alpha, \gamma \in \mathbb R$.

There is a sharp and interesting contrast in the scope of
application of our results.  Unlike the quantum Heisenberg spin
glasses, applying the Lie-Trotter expansion to the transverse S-K
model has the added benefit of allowing a natural path integral
representation, known as the Feynman-Kac representation, of
statistical quantities like the free energy.  More precisely, the
Feynman-Kac representation allows the expression of the partition
function as a measure on c\`{a}dl\`{a}g paths with state space the
classical Ising spin configurations on $N$ sites, $\{-1,+1\}^N$. For
this and other technical reasons we can only resolve the existence
of the free energy for the one quantum model.

The remainder of the paper is structured as follows.  Section
\ref{Results} introduces notation and states the main theorems of
the paper. Section \ref{Universality} presents quantum
generalizations of \cite{C-H} which allow us to prove universality
and control the fluctuations of the free energy for a class of
quantum spin glasses which include the spin-$\ffrac 12$ quantum
Heisenberg spin glasses.  In addition, we adapt the techniques of
\cite{Guerra-Ton-2} to prove exponential decay of the fluctuations
of models with Gaussian disorder. Finally, Section \ref{Pressure}
details the existence of the pressure of the transverse S-K model.

\section{Results}\label{Results}

We begin by giving a bit of notation to be used below. We consider a
collection $\xi_I$ of random variables, where $I$ is in some finite
indexed set $\mathfrak I$ and denote $\E \left[ \cdot \right]$ and
$\mathbb P \left( \cdot \right)$ integration with respect to this
collection. Examples of index sets that we have in mind include the
collection of all $r$-tuples of sites in a system of $N$ particles.
In the simplest case the index set $\mathfrak I$ consists of all
pairs $(i,j)$ (or more generally some subset of pairs). Unless
otherwise specified, the variables $\xi_I$ are assumed to be i.i.d.
according to some fixed random variable $\xi$ satisfying the
conditions
\begin{equation*}
\mathbb E\left[\xi\right]=0, \: \mathbb E\left[\xi^2\right]=1, \:
\mathbb E\left[|\xi|^3\right] < \infty.
\end{equation*}
When explicitly considering Gaussian environments we denote the
random variables by $g_I$.

Let $\underline S$ represent the $N$-tuple of Pauli vectors and
consider the Hamiltonian
\begin{equation*}
\mathcal H_N(\xi) = \sum_{I \in \mathfrak I} \xi_I X_{I}(\underline
S)
\end{equation*}
where each $X_{I}(\underline S)$ is a self-adjoint polynomial in the
spin operators, i.e. $X^*_{I}(\underline S)=X_{I}(\underline S)$. We
define the associated partition function and quenched `pressure' by
\begin{equation*}
Z_N(\beta,\xi)= \TR\left(e^{\beta \mathcal H_N}\right); \: \: \:
\alpha_N(\beta,\xi)= \E \left[ \log Z_N(\beta, \xi) \right].
\end{equation*}
Note that for convenience we have omitted the minus sign from the
expression for the Gibbs-Boltzman operator.  For any operator $A$ we
denote its operator norm by $\|A\|:= \sup_{v \in \mathbb V}
\left[\frac{\left(A v,A v\right)}{(v,v)}\right]^{\frac 12}$. With
this notation we have the following result:
\begin{theorem}\label{Uni-Fluc}
Let $\xi$ be random variable with mean $0$, variance $1$, and
$\E\left[\left|\xi\right|^3 \right] < \infty$  and let $g$ be a
standard normal random variable.  Let $\mathfrak I_N$ index the
interactions between particles at the system size $N$. Suppose
\begin{equation*}
\sum_{I \in \mathfrak I_N} \|X_I(\underline S)\|^3 = o(N).
\end{equation*}

Then for any $\beta \in \mathbb R$,
\begin{equation*}
\left|\frac{1}{N}\alpha_N(\beta,\xi) -
\frac{1}{N}\alpha_N(\beta,g)\right| = o(1)
\end{equation*}
as $N$ tends to infinity.  Moreover
\begin{equation*}
\E \left[ \left|\frac{1}{N} \log Z (\beta, \xi) -
\frac{1}{N}\alpha_N(\beta,\xi)\right|^3\right] \leq \frac{\sqrt
{|\mathfrak I_N|} o(N)}{N^3}.
\end{equation*}
\end{theorem}
\begin{remark}
In the quantum Heisenberg Hamiltonians, the norm of each summand is
bounded by $\frac{C}{\sqrt N}$ and the number of pairs is of the
order $N^2$. As a result,
\begin{equation*}
\sum_{I \in \mathfrak I_N} \|X_I(\underline S)\|^3=
O(N^{\frac{1}{2}})
\end{equation*}
and the theorem is immediately applicable.
\end{remark}

The error bounds get better if one assumes the random variable $\xi$
has higher moments.  At the extreme end, we consider the case of
fluctuations for Gaussian environments:
\begin{proposition}\label{concentration}
Let $g$ be a standard normal random variable.  Then
\begin{equation*}
\mathbb P \left( \left|\log Z_N(\beta, g) - \alpha_N(\beta,g)\right|
\geq u \right) \leq 2 e^{\left(-\frac{u^2}{\sum_{I \in \mathfrak
I_N} \beta^2 \|X_I\|^2}\right)}.
\end{equation*}
\end{proposition}

\begin{remark}
Note the generality with which these results are stated.  In the
introduction we advertised their application to mean field models.
However, choosing the operators $X_I$ and index sets $\mathfrak I_N$
appropriately allows application to a wide variety of systems.
\end{remark}

Next we take up the more subtle question of the existence of the
free energy for mean field models.  As mentioned above, we
specialize to the transverse S-K model:
\begin{equation*}
-\mathcal H_N(\lambda) = \frac{1}{2 \sqrt N}\sum_{i,j=1}^N g_{i,j}
S^{(z)}_i S^{(z)}_j + \lambda \sum_{j=1}^N S^{(x)}_j.
\end{equation*}
In light of Theorem \ref{Uni-Fluc}, we have assumed the interactions
are Gaussian.
\begin{theorem}\label{Trans-Press}
Let $\beta, \lambda > 0$ be fixed.  Then
\begin{equation*}
\lim_{N \rightarrow \infty} \frac{-1}{\beta N} \E \left[\log
\TR\left( e^{-\beta\mathcal H_N(\lambda)} \right)\right]
\end{equation*}
exists and is finite.  Moreover, there exists a $K>0$ so that the
following concentration property holds:
\begin{equation*}
\mathbb P \left( \left|\frac{-1}{N} \log \TR \left(e^{-\beta\mathcal
H_N(\lambda)}\right) + \frac{1}{N}\E \left[ \log \TR
\left(e^{-\beta\mathcal H_N(\lambda)}\right) \right] \right| \geq u
\right) \leq 2 e^{- N \frac{K u^2}{\beta^2}}.
\end{equation*}
\end{theorem}

\begin{remark}
For readability, we assume here that the disordered portion of the
Hamiltonian in \eqref{Trans-Ising} is a two body interaction.
However analogous arguments allow one to treat p-spin models where
we replace the disordered portion by
\begin{equation*}
-\mathcal H_N^{dis}(S^{(z)}) = N \sum_{r=1}^\infty
\frac{a_r}{N^{\frac{r}{2}}}\sum_{i_1, \dots i_r}g_{i_1 \dots
i_r}\prod_{k=1}^r S^{(z)}_{i_k}
\end{equation*}
Here $\{g_{i_1 \dots i_r}\}$ is a collection of independent standard
Gaussian random variables and $\sum_1^\infty a_r^2 q^{r}$ is even,
convex and continuous as a function of $q$ on $[-1,1]$.  See
\cite{A-S-S} and \cite{Guerra-Ton-2}.
\end{remark}

\section{Universality and Fluctuations}\label{Universality}
Before proving Theorem \ref{Uni-Fluc}, we adapt the methods of
\cite{C-H} so as to apply them to a wide variety of quantum spin
systems. It turns out that the line of argument given there is
robust enough to be followed in the quantum case, though the
calculations must be adjusted to accommodate the non-commutative
setting.

In the general setup, we consider a collection of self adjoint
operators $\{X_i\}_{i=1}^d$ and $\mathcal H_0$ defined on some
 finite dimensional Hilbert space. Let
\begin{align*}
\mathcal H(\xi) = \sum_{1}^d \xi_i X_i + \frac{1}{\beta} \mathcal H_0\\
Z(\beta, \xi) = \TR\left( e^{\beta \mathcal H(\xi)}\right) \\
\alpha(\beta,\xi) = \E \left[\log Z(\beta, \xi)\right]
\end{align*}
denote the Hamiltonian, partition function, and quenched `pressure'
of a system.  We define the thermal average, Duhammel two point
function, and the three point function for  the operators $A, B,$
and $C$ as follows:
\begin{align*}
\langle A \rangle  = &  \frac{\TR \left( A e^{\beta \mathcal H(\xi)} \right)}{Z(\beta, \xi)}\\
\left(A,B\right) = & \frac{\int_0^1 \TR \left( A e^{u \beta \mathcal H(\xi)}B e^{(1-u) \beta \mathcal H(\xi)} \right)\textd u}{Z(\beta, \xi)} \\
\left(A,B,C\right) = & \frac{\int_0^1 \int_0^1 u \TR \left(A e^{s u
\beta \mathcal H(\xi)}B e^{(1-s)u \beta \mathcal H(\xi)} C e^{(1-u)
\beta \mathcal H(\xi)}\right)\textd s \; \textd u }{Z(\beta, \xi)}.
\end{align*}

For the convenience of the reader, we recall a generalized
integration-by-parts lemma proved in \cite{C-H}.  The left hand side
of \eqref{IBPbound} is well known to be zero if the randomness is
Gaussian. This fact will play a role in what follows.
\begin{lemma}\label{IBP}
Let $\xi$ be a real valued random variable such that
$\E\left[|\xi|^3\right] < \infty$ and $\E \left[ \xi \right] = 0$.
Let $F: \mathbb R \rightarrow \mathbb R$ be twice continuously
differentiable with $\|F''\|_\infty = \sup_{x \in \mathbb R}
\left|F''(x)\right| < \infty.$  Then
\begin{equation}\label{IBPbound}
\left|\E \left[\xi F(\xi)\right]- \E
\left[\xi^2\right]\E\left[F'(\xi)\right]\right| \leq \frac{3}{2}
\|F''\|_\infty \E \left[ \left|\xi\right|^3 \right].
\end{equation}
\end{lemma}

We first provide an expansion of the quenched pressure:
\begin{lemma}\label{press-deriv}
\begin{equation*}
\frac{\partial \alpha(\beta, \xi)} {\partial \beta} = \beta \E
\left[ \sum_{i=1}^d \left(X_i,X_i\right) - \langle X_i \rangle^2
\right] + 9 O(\beta^2) \E\left[ \left|\xi\right|^3\right]
\left(\sum_{i=1}^d \|X_i\|^3 \right)
\end{equation*}
where $|O(\beta^2)| \leq \beta^2.$
\end{lemma}
\begin{remark}
There is no correction in this formula when $\xi$ is Gaussian since
the integration-by-parts formula is exact in this case .
\end{remark}
\begin{proof}
To begin the derivation, we have
\begin{equation*}
\frac{\partial \alpha(\beta, \xi)}{\partial \beta}=\sum_{i=1}^d \E
\left[ \xi_i \langle X_i \rangle \right].
\end{equation*}
Let us define $\mathcal H_i (z) = \sum_{j \neq i} \xi_j X_j + z X_i
+ \frac 1\beta \mathcal H_0$, denote the corresponding Gibbs state
by $\langle \cdot \rangle_i^{(z)}$ and define the function $F_i(z) =
\langle X_i \rangle_i^{(z)}$.  Then
\begin{equation*}\label{expan}
\E \left[ \xi_i \langle X_i \rangle \right] = \E \left[ \xi_i
F_i(\xi_i) \right].
\end{equation*}
To expand Equation \eqref{expan}, we calculate the first and second
derivatives of $F_i$.  Applying the Lie-Trotter expansion, i.e. that
$e^{A+B}= \lim_k \prod_1^k e^{\ffrac A k}e^{\ffrac B k}$, with $A =
\beta z X_i$ and $B = \beta \mathcal H_i(z) - \beta z X_i $, we find
that the first and second derivatives of $F_i$ take the form:
\begin{align*}
F_i'(z)=& \beta \left[\left( X_i, X_i \right)^{(z)}_i - \left(\langle X_i \rangle^{(z)}\right)^2\right]\\
F_i''(z)=& \beta^2 \left[ 2 \left( X_i, X_i, X_i \right)^{(z)} - 3
\left( X_i, X_i \right)^{(z)}\langle X_i \rangle^{(z)} + 2
\left(\langle X_i \rangle^{(z)} \right)^3 \right].
\end{align*}

In order to prove the lemma we must bound $F_i''$.  To this end we
claim that for any $a_1, \dots, a_n > 0$ so that $a_1+ \cdots +a_{n}
=1$ and any self-adjoint operators $X, \mathcal H$
\begin{equation}\label{Corr-bound}
\left|\TR\left(Xe^{a_1 \mathcal H} \cdots X e^{a_n \mathcal H}
\right)\right| \leq \|X\|^n \TR \left(e^{\mathcal H} \right)
\end{equation}
Indeed, we may assume by continuity that $a_j=\frac{k_j}{2^m}$ for
some $m>0$ and some sequence of positive integers $(k_j)$ summing to
$2^m$.

For any sequence of operators $(B_j)_{j=1}^{2k}$,
\begin{equation}\label{R-P}
\left|\TR\left(\prod_{j=1}^{2k} B_j\right)\right| \leq
\prod_{j=1}^{2k} \TR\left(\left[B_j
B_j^*\right]^{k}\right)^{\frac{1}{2k}}.
\end{equation}
This can be seen as a specific case of the general method of
chessboard estimates, Theorem $4.1$ of \cite{FILS}. Applying
\eqref{R-P} to the left hand side of \eqref{Corr-bound} with $B_j
\in \{X e^{\frac{\mathcal H}{2^m}}, e^{\frac{\mathcal H}{2^m}}\}$ we
have
\begin{equation}\label{interstep}
\left|\TR\left(Xe^{a_1 \mathcal H} \cdots X e^{a_n \mathcal H}
\right)\right| \leq \TR\left(\left[ Xe^{\frac{\mathcal H}{2^{m-1}}}X
\right]^{2^{m-1}}\right)^{\frac{n}{2^{m}}}  \TR  \left(e^{\mathcal
H} \right)^{\frac{2^{m}-n}{2^{m}}}
\end{equation}
Another application of \eqref{R-P} implies
\begin{equation*}
\TR\left( \left[Xe^{\frac{\mathcal H}{2^{m-1}}}X
\right]^{2^{m-1}}\right) \leq \TR\left(
X^{2^{m+1}}\right)^{\frac{1}{2}} \TR \left(e^{2\mathcal H}
\right)^{\frac {1}{2}}.
\end{equation*}
Using this bound on the right hand side of \eqref{interstep} and
letting $m$ pass to infinity proves the bound \eqref{Corr-bound}.

It follows that $\|F_i''\|_\infty \leq 6 \beta^2 \|X_i\|^3.$
Recalling that $\E \left[\xi^2\right] =1$, Lemma \ref{IBP} in
conjunction with the previous calculations imply
\begin{equation*}
\left|\frac{\partial \alpha(\beta, \xi)} {\partial \beta} - \beta \E
\left[ \sum_{i=1}^d \left(X_i,X_i\right) - \langle X_i \rangle^2
\right] \right| \leq 9 \beta^2 \E\left[ |\xi|^3\right]
\left(\sum_{i=1}^d \|X_i\|^3 \right).
\end{equation*}
\end{proof}

Next we use this expansion to compare the quenched `pressure' for
$\xi$ to that of a Gaussian environment $g$ where $g$ is a standard
normal.
\begin{lemma}\label{comp}
Let $\{\xi_i\}$ and $\{g_i\}$ be collections of i.i.d random
variables distributed according to $\xi$ and $g$ respectively. For
any $\beta \in \mathbb R$,
\begin{equation*}
|\alpha(\beta,\xi) - \alpha(\beta,g)| \leq 9 |\beta|^3 \E\left[
|\xi|^3\right] \left(\sum_{i=1}^d \|X_i\|^3 \right)
\end{equation*}
\end{lemma}
\begin{proof}
The similarity between the proof of this lemma and that of
Proposition 7 \cite{C-H} means that we will be extremely brief.
Consider the interpolating partition function and corresponding
quenched `pressure' defined by
\begin{align*}
Z(s, t-s) = & e^{\sqrt{s} \left(\sum_{i=1}^d \xi_i X_i\right) +
\sqrt{t-s}\left(\sum_{i=1}^d g_i X_i\right) + \mathcal H_0} \\
\alpha^{(t)}(s) = & \E \log Z(s,t-s)
\end{align*}
respectively.  We have
\begin{equation*}
\alpha^{(t)}(t)= \alpha (\sqrt t, \xi); \: \: \: \alpha^{(t)}(0)=
\alpha (\sqrt t, g).
\end{equation*}
Lemma \ref{press-deriv} along with independence between the two
environments implies that for all $s \in [0,t]$ we have
\begin{equation*}
\left| \frac{\partial \alpha^{(t)}(s)}{\partial s} \right| \leq 9
\sqrt{t} \E\left[ |\xi|^3\right] \left(\sum_{i=1}^d \|X_i\|^3
\right).
\end{equation*}
For $\beta \geq 0$, if we let $t=\beta^2$ and integrate this
inequality the result follows.  For $\beta < 0$ we instead consider
the environments $-\xi, -g$.
\end{proof}

Next we attend to the fluctuations of the `pressure' determined by
the random environment $(\xi_i)$:
\begin{lemma}\label{xi-fluc}
There exists some universal constant $c>0$ so that
\begin{equation*}
\E \left[ |\log Z (\beta, \xi) - \alpha (\beta,\xi)|^3\right] \leq c
\; \E \left[ |\xi|^3 \right] \beta^3 \sqrt d \left( \sum_{i=1}^d
\|X_i\|^3 \right).
\end{equation*}
\end{lemma}
\begin{proof}
Consider the filtration $\mathcal F_k= \sigma \{\xi_1 \dots \xi_k\},
k \geq 1$ determined by the sequence of independent random variables
$(\xi_k)$.  Let
\begin{equation*}
\Delta_i:= \E \left[\log Z(\beta,\xi)\big| \mathcal F_i\right] - \E
\left[\log Z(\beta,\xi)\big| \mathcal F_{i-1}\right]
\end{equation*}
 We have
\begin{equation*}
\log Z(\beta, \xi)- \alpha(\beta,\xi)= \sum_{i=1}^d \Delta_i.
\end{equation*}
Burkholder's martingale inequality implies the existence of a
universal constant $c'$ so that
\begin{equation*}
\E \left|\sum_{i=1}^d \Delta_i\right|^3 \leq c' \E
\left(\sum_{i=1}^d \Delta_i^2\right)^{\frac{3}{2}}.
\end{equation*}
To bound the increment $\Delta_i$, consider the partition function
\begin{equation*}
Z_i(\beta, \xi) = \TR \left(e^{\beta \psi_i}\right)
\end{equation*}
where $\psi_i= \psi_i(\xi) := \mathcal H (\xi) - \xi_i X_i$.  Since
$Z_i(\beta, \xi)$ is independent of $\xi_i$,
\begin{equation*}
\Delta_i = \E \left[\log \frac{Z(\beta,\xi)}{Z_i(\beta, \xi)}\big|
\mathcal F_i\right] -  \E \left[\log \frac{Z(\beta,\xi)}{Z_i(\beta,
\xi)}\big| \mathcal F_{i-1}\right].
\end{equation*}
We use this identity to estimate $\Delta_i$.

We claim that
\begin{equation*}
\frac{Z(\beta,\xi)}{Z_i(\beta, \xi)} \leq e^{\beta |\xi_i| \cdot
\left \|X_i\right \|}.
\end{equation*}
Indeed, the Lie-Trotter formula implies
\begin{equation*}
Z(\beta,\xi) = \lim_{k \rightarrow \infty}
\TR\left(\prod_{j=1}^{2^k} e^{\frac{\beta
\psi_i}{2^k}}e^{\frac{\beta \xi_i X_i}{2^k}}\right)=\lim_{k
\rightarrow \infty} \TR\left(\prod_{j=1}^{2^k} e^{\frac{\beta \xi_i
X_i}{2^{k+1}}}e^{\frac{\beta \psi_i}{2^{k}}}e^{\frac{\beta \xi_i
X_i}{2^{k+1}}}\right) .
\end{equation*}
Since
\begin{equation*}
\left \| e^{\frac{\beta \xi_i X_i}{2^k}}\right \| \leq
e^{\frac{\beta |\xi_i| \cdot \| X_i\|}{2^k}},
\end{equation*}
Inequality \eqref{Int-claim} applied to the righthand side for $k$
finite gives
\begin{equation*}
\TR\left(\prod_{j=1}^{2^k} e^{\frac{\beta
\psi_i}{2^k}}e^{\frac{\beta \xi_i X_i}{2^k}}\right) \leq e^{\beta
|\xi_i| \cdot \| X_i\|}\TR\left(e^{\beta \psi_i}\right)
\end{equation*}
from which our claim follows.

From this we estimate:
\begin{equation*}
|\Delta_i| \leq \beta \|X_i\| (|\xi_i| + \E |\xi_i|).
\end{equation*}
Therefore,
\begin{align*}
\E |\log Z(\beta,\xi) - \alpha(\beta,\xi)|^3 \leq c' \E
\left(\sum_{i=1}^d \Delta_i^2\right)^{\frac{3}{2}} \leq & c' \beta^3
\left(\sum_{i=1}^d \|X_i\|^2 \E (|\xi_i| + \E |\xi_i|)^2\right)^{\frac{3}{2}}\\
\leq & c \beta^3\E |\xi_i|^{3} \sqrt{d} \left(\sum_{i=1}^d \|X_i\|^3
\right)
\end{align*}
\end{proof}

\begin{proof}[Proof of Theorem \ref{Uni-Fluc}]
Our first theorem now follows as an application of the above
machinery: the first statement is an application of Lemma
\ref{comp}, while the second follows from Lemma \ref{xi-fluc}.
\end{proof}

Finally, we consider fluctuations in Gaussian environments:
\begin{proof}[Proof of Proposition \ref{concentration}]
Let $Z_\beta(t)$ be defined as the auxiliary partition function
given by two independent collections of Gaussian disorder $g^{(1)}$
and $g^{(2)}$,
\begin{equation*}
Z_\beta(t)= \TR\left(e^{\beta \sqrt{t}\sum_{i=1}^d g^{(1)}_i X_i+
\beta \sqrt{1-t}\sum_{i=1}^d g^{(2)}_i X_i + \mathcal H_0}\right)
\end{equation*}
with $t$ an interpolation parameter varying between $0$ and $1$. Let
$\E_j$ denote the average with respect the random variables
$g^{(j)}$ for $j=1,2$. Given any $s \in \mathbb R$, let
\begin{equation*}
Y(t)=\exp (s \E_2 \log Z_\beta(t)); \: \: \: \phi(t) = \log \E_1
\left[Y(t)\right].
\end{equation*}
We note that
\begin{equation}\label{integral}
\phi_N(1)-\phi_N(0) = \log \E \left[\exp( s (\log Z(\beta, g) -
\alpha(\beta,g)))\right].
\end{equation}

In order to estimate this difference, consider
\begin{equation*}
\phi'(t)= \frac{s}{2 \E_1 \left[Y(t)\right]} \E_1 \left[ Y(t) \E_2
\left[\frac{1}{\sqrt{t}}\left \langle \sum_{i=1}^d g^{(1)}_i X_i
\right \rangle_t - \frac{1}{\sqrt{1-t}}\left \langle \sum_{i=1}^d
g^{(2)}_i X_i \right \rangle_t \right] \right]
\end{equation*}
where the notation $\langle \cdot \rangle_t$ represents the Gibbs
state induced by the interpolating Hamiltonian.  A calculation
involving the Lie-Trotter expansion and the Gaussian version of the
integration-by-parts formula implies
\begin{equation*}
\phi'(t) =\frac{s^2\beta^2}{2 \E_1 \left[Y(t)\right]} \E_1
\left[Y(t) \sum_{i=1}^d \langle X_i \rangle_t^2 \right]
\end{equation*}
so that
\begin{equation}\label{gauss-deriv}
|\phi'(t)| \leq \frac{s^2 \beta^2 \sum_{i=1}^d \|X_i \|^2}{2}.
\end{equation}
Using Equation \eqref{gauss-deriv} and the fact that $e^{|x|} \leq
e^{x} + e^{-x}$ we have
\begin{equation*}
\exp(|s||\log Z(\beta, g)- \alpha(\beta, g)|) \leq 2 e^{\frac{s^2
\beta^2 \sum_{i=1}^d \|X_i \|^2}{2}}.
\end{equation*}

Finally, applying Markov's inequality we have
\begin{equation*}
\mathbb P \left( |\log Z(\beta, g) - \alpha(\beta, g)| \geq u
\right) \leq 2 e^{\frac{s^2 \beta^2 \sum_{i=1}^d \|X_i \|^2}{2}-s u}
\end{equation*}
for any $s \in \mathbb R$.  Optimizing over $s$ concludes the lemma.
\end{proof}

\section{Existence of the Pressure}\label{Pressure}
Recalling the $\mathfrak {su}(2)$ formalism, we choose a preferred
basis for $\mathbb V_N$ consisting of tensor products of
eigenvectors for the operators $\{S^{(z)}_j\}$. Denoting the
eigenvector for $S^{(z)}$ which corresponds to the eigenvalue $+1$
by $\ket{+}$ and the eigenvector for which corresponds to the
eigenvalue $-1$ by $\ket{-}$, we may identify this preferred basis
with classical Ising spin configurations $\sigma \in \{-1,+1\}^N$.
For each $\sigma$, we denote the corresponding basis vector by
$\ket{\sigma}$.

The proof of Theorem \ref{Trans-Press} proceeds in two steps. The
first step consists of a concentration estimate following
essentially the same argument as that of Proposition
\ref{concentration}. Widening the scope beyond the transverse Ising
spin glass, for this first step we consider quantum Hamiltonians of
the form
\begin{equation}\label{F-K-Ham}
\mathcal H_N := \mathcal{H}_N^{\text{dis}}(S^{(z)}) +
\mathcal{H}_N^{\text{det}}
\end{equation}
We assume that only $\mathcal{H}_N^{\text{dis}}$ involves Gaussian
disorder and that the deterministic operator
$\mathcal{H}_N^{\text{det}}$ takes a sufficiently nice form so as to
admit a Feynman-Kac representation in terms of the basis of
eigenvectors for the $S^{(z)}$ operators. By this we mean that
\begin{equation*}
\bra{\sigma} \exp(-u\mathcal{H}_N^{\text{det}}) \ket{\tilde{\sigma}}
\geq 0
\end{equation*}
for all $u \geq 0$ and all spin configurations $\sigma,
\tilde{\sigma}$.  Assuming that all off diagonal matrix elements of
$\mathcal{H}_N^{\text{det}}$ are negative gives a necessary and
sufficient condition which guarantees the existence of a Feynman-Kac
representation.  As mentioned in Section \ref{Results}, for our
treatment $\mathcal{H}_N^{\text{dis}}$ takes the form
\begin{equation*}
-\mathcal H_N^{\text{dis}}(S^{(z)}) = \frac{1}{2 \sqrt N}\sum_{i,
j=1}^N g_{i j} S^{(z)}_{i}S^{(z)}_{j}.
\end{equation*}
where the collection $\{g_{i,j}\}$ is assumed to be i.i.d. standard
normal, though more general interactions involving the
$\{S_j^{(z)}\}$ may be considered.

To illustrate the significance of the Feynman-Kac representation,
recall that by adding a suitably chosen diagonal matrix to $\mathcal
H_N^{\text{det}}$ we can force the rows of the matrix representation
of $\mathcal H_N^{\text{det}}$ to sum to $0$, which implies that the
(modified) Hamiltonian $\mathcal H_N^{\text{det}}$ is the generator
of a continuous time Markov chain with state space $\{-1,+1\}^N$.

Consider any matrix element $\bra{\sigma} e^{-u \mathcal H_N}
\ket{\tilde{\sigma}}$.  Expanding the exponential for finite $k$
using the Lie-Trotter formula with $A = -u \mathcal
H_N^{\text{dis}}(S^{(z)})$ and $B= -u \mathcal H_N^{\text{det}}$ and
inserting the complete orthogonal set $\{\ket{\sigma'}\}$ between
each factor $e^{\frac{A}{k}}e^{\frac{B}{k}}$ and then passing to the
limit in $k$, it is not difficult to see that
\begin{equation*}
\bra{\sigma} e^{-u \mathcal H_N} \ket{\tilde{\sigma}} =
\int_{\sigma(0)=\sigma} e^{\int_0^u - \mathcal
H_N^{\text{dis}}(\sigma(u)) \textd u} \1_{\sigma(u)=\tilde{\sigma}}
\textd \nu(\sigma)
\end{equation*}
where $\nu$ is the induced Markov chain measure and
\begin{equation*}
\mathcal H_N^{\text{dis}}(\sigma) = \bra{\sigma}\mathcal
H_N^{\text{dis}}(S^{(z)}) \ket{\sigma}
\end{equation*}.

Moreover, we can view the augmentation by the diagonal matrix as
introducing a weighting to each spin configuration which corresponds
to the amount of time the process spends in each state along its
trajectory. Thus, the original Hamiltonian yields an un-normalized
measure on paths taking values in $\{-1,+1\}^N$. Let us also note,
as is the case in the transverse S-K model, that the diagonal matrix
is constant.

Returning to our original Hamiltonian \eqref{F-K-Ham}, the upshot is
that via the Feynman-Kac transformation we may represent the
partition function $Z_N$ associated to the Hamiltonian $\mathcal
H_N$ by
\begin{equation*}
Z_N = \TR \left(e^{-\beta \mathcal H_N}\right) =
\int_{\Omega}e^{\int_0^\beta \frac{1}{2\sqrt N} \sum_{i,j=1}^N
g_{ij}\sigma_i(u) \sigma_j(u) \textd u} \textd
\nu{\left(\underline{\sigma}\right)}
\end{equation*}
where $\nu$ is a measure on spin configuration paths determined by
$\mathcal H^{\text{det}}_N$ and $\Omega$ is the space of
c\`{a}dl\`{a}g paths taking values in $[-1,1]$. We use this
representation implicitly throughout Section \ref{Pressure}.  In the
case of the transverse Ising model, one may check that the induced
measure is in fact (proportional to) the Markov chain measure
defined by starting from an initial configuration and evolving in
time via a spin flip process determined by flipping the spin value
at each site according to the arrivals of independent Poisson
processes of rate $\lambda$. In order to prove convergence of the
quenched pressure we shall need to consider restricted partition
functions $Z^A_N$ defined via the Feynman-Kac transformation by
\begin{equation*}
\frac{1}{N} \log Z^A_N = \frac{1}{N}\log \int_{A}e^{\int_0^\beta
\frac{1}{2\sqrt{N}} \sum_{i,j=1}^N g_{ij}\sigma_i(u) \sigma_j(u)
\textd u} \textd \nu{(\underline{\sigma})}
\end{equation*}
where $A=A_N$ is some deterministic (i.e. not depending on Gaussian
disorder) subset of $\Omega$.

For any pair of spin configurations $\sigma$ and $\tilde{\sigma}$,
let
\begin{equation*}
R(\sigma, \tilde{\sigma})= \frac{1}{N}\sum_{i=1}^N \sigma_i \tilde
\sigma_i,
\end{equation*}
$\underline{\sigma}$ denote the spin path $\{\sigma(u) : u \in
\left[0,\beta\right]\}$ and
\begin{equation*}
\zeta_N(\underline{\sigma})= \int_0^\beta \frac{1}{2\sqrt{N}}
\sum_{i,j=1}^N g_{ij}\sigma_i(u) \sigma_j(u) \textd u.
\end{equation*} denote the Gaussian random variable defined by the
classical Gibbs weight of the spin path $\underline{\sigma}$.  An
easy calculation shows that
\begin{equation*}
\E\left[ \zeta_N(\underline{\sigma}) \zeta_N( \underline{\tilde
\sigma}) \right]= \frac{N}{4} \int_0^\beta \int_0^\beta
R^2(\sigma(u), \tilde{\sigma}(s)) \textd u \textd s
\end{equation*}
where the function $R$ is the classical overlap for Ising spin
configurations.

\begin{lemma}\label{conc2} Let $\beta>0$ be given.  Fix any Feynman-Kac representable deterministic Hamiltonian $\mathcal
H_N^{\text{det}}$.  Then
\begin{equation*}
\BbbP \left( \left|\frac{1}{N} \log Z^A_N - \frac{1}{N}\E \left[
\log Z^A_N \right]\right| \geq u \right) \leq 2 \exp\left(- \frac{2
u^2}{\beta^2}N\right).
\end{equation*}
\end{lemma}
\begin{proof}
With the appropriate modifications, the method of proof of
Proposition \ref{concentration} may be followed here:  Let
$Z^A_N(t)$ be defined as the auxiliary partition function given by
two independent collections of Gaussian disorder $g^{(1)}$ and
$g^{(2)}$,
\begin{equation*}
Z^A_N(t)= \int_{A}e^{\sqrt{t}\zeta^{(1)}_N(\underline{\sigma})+
\sqrt{1-t}\zeta^{(2)}_N(\underline{\sigma})} \textd
\nu(\underline{\sigma})
\end{equation*}
with $t$ an interpolation parameter varying between $0$ and $1$ and
$\zeta^{(i)}$ the Gaussian corresponding to $g^{(i)}$. Let $Y_N(t)$
and $\phi_N(t)$ be defined in terms of $Z^A_N(t)$ as in
\eqref{exp-mom1} and \eqref{exp-mom2}.

Replacing the corresponding quantities appearing in the proof of
Proposition \ref{concentration} we have, using the Feyman-Kac
representation,
\begin{multline*}
\phi_N'(t) =\frac{s^2}{2 \E_1 \left[Y_N(t)\right]}\\
\E_1 \left[Y_N(t) \int_{A_N \times A_N} \textd
\nu(\underline{\sigma}) \textd \nu (\underline{\tilde{\sigma}})
\left(\frac{N}{4} \int_0^\beta \int_0^\beta R^2(\sigma(u),
\tilde{\sigma}(s)) \textd u \textd s\right) \: w_N(t,
\underline{\sigma}) w_N(t, \underline{\tilde{\sigma}})\right]
\end{multline*}
where $w_N(t, \underline{\sigma})$ is the truncated `Gibbs weight'
corresponding to the event $A_N$.  Thus
\begin{equation*}
\left|\phi_N'(t)\right| \leq \frac{s^2 \beta^2 N}{8}.
\end{equation*}
The bound now follows as in the proof of Proposition
\ref{concentration}.

\end{proof}

\begin{remark}
More generally, the p-spin models may also be treated via the method
employed here, though the bound stated in the lemma must be modified
slightly.
\end{remark}

Unfortunately the use of the Feynman-Kac transformation alone does
not allow our method to go through.  In particular, we were unable
to treat the quantum system with deterministic quadratic couplings
in the $x$ and $y$ directions:  the ferromagnetic version does
permit a Feynman-Kac representation but the interaction is convex,
which turns out to have exactly the wrong sign in the expression for
the derivative of the interpolating `pressure'.

The only natural example that we found amenable to our method is the
transverse field Ising model.  Notice that the deterministic portion
of this particular Hamiltonian is \textit{linear}, which simplifies
the interpolation scheme that we employ to analyze the
thermodynamics at different system sizes.  For inverse temperature
$\beta$ and transverse field strength $\lambda>0$, we refer to the
partition function of this model by $Z_N(\beta,\lambda)$ and denote
$p_N(\beta,\lambda)=-\frac{1}{N} \log Z_N(\beta,\lambda)$.

\begin{lemma}\label{quenched-pressure}
Let $\beta, \lambda > 0$ be fixed.  Then
\begin{equation*}
\lim_{N \rightarrow \infty} \E \left[p_N(\beta , \lambda)\right]
\equiv p(\beta, \lambda)
\end{equation*}
exists.
\end{lemma}

\begin{proof}
This proof, like that of the previous lemma, relies on the proof of
an analogous statement in \cite{Guerra-Ton-2}. The main idea is to
partition the space of paths into subsets on which we may control
the time correlated self overlap $R(\sigma(u),\sigma(s))$.

To this end, let $\epsilon, \; \delta > 0$ be fixed, where for
convenience we assume $\frac{\beta}{\delta} \in \mathbb N$.  For any
function $g:[0,\beta]\times [0,\beta] \rightarrow [-1,1]$ we define
the event $A_g(\delta, \epsilon)$ by
\begin{multline*}
A_g(\delta, \epsilon)= \{\sigma : g(i\delta,j\delta)\leq
R(\sigma(i\delta),
\sigma(j \delta)) < g(i\delta, j \delta) + \epsilon\; \forall \; i,j \leq \frac{\beta}{\delta}-1, \\
\: |g(u,s)-R(\sigma(u), \sigma(s))| \leq 2 \epsilon \; \forall \;
(u,s)\; \in \;[0,\beta)\times [0,\beta) \}.
\end{multline*}
Let $S_\delta(\epsilon)$ be the set of functions which are constant
on $[j\delta,(j+1)\delta)\times [k \delta, (k+1) \delta)$ for $j, k
\in \{0, \dots \frac{\beta}{\delta} -1\}$ and take values in
$\{i\epsilon: i \in [-\frac{1}{\epsilon}, \frac{1}{\epsilon}]\cap
\mathbb N\}$. We define the event
\begin{equation*}
A= A(\delta,\epsilon) = \cup_{g \in S_\delta(\epsilon)} A_g(\delta,
\epsilon).
\end{equation*}

Observe that though $A$ definitely does not cover the full sample
space $\Omega$, it is enough to prove convergence of the truncated
pressure
\begin{equation*}
p^{A}_N(\beta, \lambda) = -\frac{1}{N} \log \int_{A}
e^{\zeta_N(\underline{\sigma})}\textd \nu(\underline{\sigma}).
\end{equation*}
More precisely suppose $\epsilon= \epsilon_N, \delta = \delta_N$. We
claim that if $- \epsilon_N \log \delta_N$ is sufficiently large as
$N \rightarrow \infty$
\begin{equation*}\label{press-conv}
\lim_{N \rightarrow \infty} \E\left[ p_N(\beta,
\lambda)-p^{A}_N(\beta, \lambda)\right] = 0.
\end{equation*}

Let $\Delta \mathcal N_{i\delta}(\underline{\sigma})$ denote the
total number of jumps made by the spin path $\underline{\sigma}$ in
the time interval $[i\delta, (i+1)\delta]$.  To determine how large
to take $- \epsilon \log \delta$ with $N$, let $A_* =
\{\underline{\sigma} \text{ : } \max_{i \leq
\frac{\beta}{\delta}-1}\Delta \mathcal
N_{i\delta}(\underline{\sigma})\geq \frac{\epsilon N}{4}\}$.   Then
$A^c \subset A_*$ so that
\begin{equation*}
\frac{\int_{A^c} \exp\left(\zeta_N(\underline{\sigma})\right) \textd
\nu(\underline{\sigma})}{\int_{A}
\exp\left(\zeta_N(\underline{\sigma})\right) \textd
\nu(\underline{\sigma})} \leq \frac{\int_{A_*}
\exp\left(\zeta_N(\underline{\sigma})\right) \textd
\nu(\underline{\sigma})}{\int_{A_*^c}
\exp\left(\zeta_N(\underline{\sigma})\right) \textd
\nu(\underline{\sigma})}
\end{equation*}

By Jensen's inequality,
\begin{multline}\label{truncated-pressure}
\E \left[\log\left(1+ \frac{\int_{A_*}
\exp\left(\zeta_N(\underline{\sigma})\right) \textd
\nu(\underline{\sigma})}{\int_{A_*^c}
\exp\left(\zeta_N(\underline{\sigma})\right) \textd
\nu(\underline{\sigma})}\right) \right] \leq \\
\log\left(1+ \int_{A_*}\E\left[\frac{
\exp\left(\zeta_N(\underline{\sigma})\right) \textd
\nu(\underline{\sigma})}{\int_{A_*^c}
\exp\left(\zeta_N(\underline{\sigma})\right) \textd
\nu(\underline{\sigma})} \right]\textd
\nu(\underline{\sigma})\right).
\end{multline}
By the Cauchy-Schwarz inequality,
\begin{multline*}
\E\left[\frac{
\exp\left(\zeta_N(\underline{\sigma})\right)}{\int_{A_*^c}
\exp\left(\zeta_N(\underline{\sigma})\right)
\textd \nu(\underline{\sigma})} \right] \\
\leq \E\left[
\exp\left(2\zeta_N(\underline{\sigma})\right)\right]^{\frac
12}\E\left[ \frac{1}{\left(\int_{A_*^c}
\exp\left(\zeta_N(\underline{\sigma})\right) \textd
\nu(\underline{\sigma})\right)^2}
\right]^{\frac 12} \\
\leq \E\left[
\exp\left(2\zeta_N(\underline{\sigma})\right)\right]^{\frac
12}\E\left[\exp\left(
\frac{2}{\nu(A^c_*)}\int_{A_*^c}\zeta_N(\underline{\sigma}) \;
\textd \nu(\underline{\sigma})\right)\right]^{\frac
12}\nu(A^c_*)^{-1}.
\end{multline*}
The last line follows from Jensen's inequality applied with respect
to the path measure $\frac{\1_{A^c_*}\textd \nu}{\nu (A^c_*)}$.
Since $2\zeta_N(\underline{\sigma})$ is a Gaussian random variable
with variance $N \int_0^\beta \int_0^\beta R^2(\sigma(u), \sigma(s))
\textd u \textd s$,
\begin{equation*}
\E\left[ \exp\left(2 \zeta_N(\underline{\sigma})\right)\right] =
\exp\left(\frac{N}{2} \int_0^\beta \int_0^\beta R^2(\sigma(u),
\sigma(s))\textd u \textd s \right).
\end{equation*}
Similarly, after a short calculation we have
\begin{multline*}
\E\left[\exp\left(
\frac{2}{\nu(A^c_*)}\int_{A_*^c}\zeta_N(\underline{\sigma}) \;
\textd \nu(\underline{\sigma})\right)\right]
= \\
\exp\left(\frac{N}{2} \int_{A^c_* \times A^c_*} \int_0^\beta
\int_0^\beta R^2(\sigma(u), \tilde \sigma(s))\textd u \textd s
\frac{\textd \nu(\underline{\sigma}) \textd \nu(\underline{\tilde \sigma})}{\nu(A^c_*)^2}
\right)
\end{multline*}

As a result of the preceding estimates
\begin{equation*}
\int_{A_*}\E\left[\frac{ \exp\left(\zeta_N(\underline{\sigma})
\right) \textd \nu(\underline{\sigma})}{\int_{A_*^c}
\exp\left(\zeta_N(\underline{\sigma})\right) \textd
\nu(\underline{\sigma})} \right]\textd \nu(\underline{\sigma}) \leq
e^{\frac{N\beta^2}{2}} \frac{\nu(A_*)}{\nu(A_*^c)}.
\end{equation*}
Standard calculations imply that for all $\epsilon$ small enough and
$\delta< \epsilon^2$,
\begin{equation}\label{arrival-bound}
\frac{\nu(A_*)}{\nu(A_*^c)} \leq \frac{1}{\delta}e^{C N \epsilon
\log \delta}
\end{equation}
for some universal constant $C> 0$.  Requiring $- C \epsilon \log
\delta \geq \beta^2 + -\frac{1}{N} \log \delta$, putting estimates
together and taking the appropriate limits proves our claim.  Note
that these conditions can be arranged, for example, by letting
$\epsilon_N = N^{-\ffrac 14}$ and $\delta_N = e^{- N^{\ffrac 12}}$
and taking $N$ large.

Thus we are reduced to showing that the mean of the truncated
pressure $\E\left[ p_N^A(\beta, \lambda)\right]$ converges. For
convenience of exposition we first consider subsequences of the form
$N_k= N_0n^k$ for some $N_0, n \in \mathbb N$.  For any $k$, we may
view $\underline{\sigma} \in \Omega_{N_k}$ as an $n$-tuple of spin
paths $(\underline{\sigma}^{(1)}, \dots,\underline{\sigma}^{(n)})$
so that $\underline{\sigma}^{(l)} \in \Omega_{N_{k-1}}$.  In order
to compare thermodynamics at consecutive system sizes, let us define
the interpolating Hamiltonian
\begin{equation*}
\zeta_{N_k}(t, \underline{\sigma}) =
\sqrt{t}\zeta_{N_k}(\underline{\sigma}) + \sqrt{1-t}\sum_{l=1}^n
\zeta_{N_{k-1}}^{(l)}(\underline{\sigma}^{(l)})
\end{equation*}
where $\zeta_{N_{k-1}}^{(l)}(\underline{\sigma}^{(l)})$ involve
disorder couplings which are mutually independent and independent
from the couplings in $\zeta_{N_k}(\underline{\sigma})$ . In
addition, we introduce the partition function
\begin{equation*}
Z_N^A(t) = \int_A \exp\left(\zeta_N(t,\underline{\sigma}) \right)
\end{equation*}

Let
\begin{equation*}
\tilde{A}_g= \tilde{A}_{N_k}(\delta, \epsilon, g)=
\{\underline{\sigma} =(\underline{\sigma}^{(1)}, \dots,
\underline{\sigma}^{(n)}) \; : \; \underline{\sigma}^{(l)} \in
A_{N_{k-1}}(\delta, \epsilon, g)\}
\end{equation*}
Obviously $\tilde{A}_g \subset A_g$. Therefore
\begin{equation}\label{A-trunc}
-\log Z^{A_g}_{N_k}(t) \leq -\log Z^{\tilde A_g}_{N_k}(t)
\end{equation}

For any $g\in S_\delta(\epsilon)$ consider
\begin{equation*}
\phi^{(g)}_{N_k}(t) = -\frac{1}{N_k} \E \log Z^{\tilde
A_g}_{N_k}(t).
\end{equation*}
After bit of work employing the integration-by-parts formula we
arrive at
\begin{multline*}
\frac{\textd}{\textd t}\phi^{(g)}_{N_k}(t)= - \frac18 \E\left[ \left
\langle \int_0^\beta \int_0^\beta R^2(\sigma(u), \sigma(s)) \textd u
\textd s - \frac 1n \sum_1^n \int_0^\beta \int_0^\beta
R^2(\sigma^{(l)}(u), \sigma(s)) \textd u
\textd s \right \rangle_t^{\tilde A_g, {(1)}}\right]\\
+ \frac 18 \E \left[\left \langle \int_0^\beta \int_0^\beta
R^2(\sigma(u), \tilde{\sigma}(s)) \textd u \textd s - \frac 1n
\sum_1^n \int_0^\beta \int_0^\beta R^2(\sigma^{(l)}(u), \tilde
\sigma^{(l)}(s)) \right \rangle_t^{\tilde A_g,{(2)}} \right]
\end{multline*}
where $\langle \cdot \rangle_t^{\tilde{A}_g,{(1)}}$ corresponds to
the truncated Gibbs weight determined by the Hamiltonian
$\zeta_{N_k}(t, \underline{\sigma})$ and $\langle \cdot
\rangle_t^{\tilde{A}_g,{(2)}}$ corresponds to the product Gibbs
weight determined by an independent pair of spin paths
$\underline{\sigma}, \underline{\tilde{\sigma}}$ and Hamiltonian
$\zeta_{N_k}(t, \underline{\sigma}) + \zeta_{N_k}(t,
\underline{\tilde{\sigma}})$.  We stress that corresponding terms in
this pair Hamiltonian involve the same realizations of disorder.
From the definition of $\tilde A_g$ the first term can be bounded by
$\beta^2 \epsilon$ in absolute value. As $f$ is convex, the latter
term is less than or equal to zero.  Thus, by evaluating
$\phi^{(g)}_{N_k}$ at zero and one and applying \eqref{A-trunc} we
have
\begin{equation*}
\frac{-1}{N_k} \E \left[\log Z^{A_g}_{N_k}\right]\leq
\frac{-1}{N_{k-1}} \E \left[\log Z^{A_g}_{N_{k-1}} \right] + \beta^2
\epsilon
\end{equation*}
for any $g \in S_\delta(\epsilon)$.

Next, by Lemma \ref{conc2}, we have
\begin{equation*}
\BbbP \left( \left|\frac{1}{N_k} \log Z^{A_g}_{N_k}(\beta, \lambda)
- \frac{1}{N_k}\E \left[ \log Z^{A_g}_{N_k}(\beta, \lambda)\right]
\right| \geq u \right) \leq 2 \exp\left(-\frac{ 2 u^2}{\beta^2}N_k
\right).
\end{equation*}
Setting $u= \epsilon$, there exists a set $\mathcal S$ and a
universal constant $D>0$ so that
\begin{equation*}
\BbbP(\mathcal S) \geq 1- \frac{4}{ \delta^2 \epsilon}\exp\left(-
\frac{2\epsilon^2}{\beta^2 n} N_k\right)
\end{equation*}
and so that on $\mathcal S$
\begin{equation*}
\frac{-1}{N_k} \log Z^{A_g}_{N_k}(\beta, \lambda) \leq
\frac{-1}{N_{k-1}} \log Z^{A_g}_{N_{k-1}}(\beta, \lambda) + D
\beta^2 \epsilon
\end{equation*}
for all $g \in S_{\delta}(\epsilon)$.

Now let us choose $\delta = \exp(- \frac{1}{2(\beta\epsilon)^2})$
and $\epsilon = N_k^{-\ffrac 14}$. Then a standard application of
the Borel-Cantelli lemma implies that there exists $D>0$ so that
\begin{equation*}
\BbbP \left( \frac{-1}{N_k} \log Z^{A_g}_{N_k}(\beta, \lambda) \geq
\frac{-1}{N_{k-1}} \log Z^{A_g}_{N_{k-1}}(\beta, \lambda) + D\beta^2
N_k^{-\ffrac 14} \text{ for some $g \in
S_{\delta(N_k)}(N_k^{-\ffrac14})$ } i.o. \right) =0.
\end{equation*}

Therefore, for $N_k$ large enough (with the choice of $\epsilon$ and
$\delta$ depending on $N_k$ as above) we have
\begin{multline*}
Z^{A_{N_k}}_{N_k}(\beta, \lambda) = \sum_{g \in
S_\delta(\epsilon)} Z^{A(\delta, \epsilon, g)}_{N_k}(\beta, \lambda)\\
\geq \sum_{g \in S_\delta(\epsilon)} e^{-D \beta^2 N_k^{\ffrac
34}}\left[Z^{A(\delta, \epsilon, g)}_{N_{k-1}}(\beta, \lambda)\right]^n\\
\geq e^{-D \beta^2 N_k^{\ffrac 34}}e^{-(n-1)\beta^{-2} N_k^{\ffrac
12}}N_k^{\frac{1-n}{4}}\left[Z^{A_{N_{k}}}_{N_{k-1}}(\beta,
\lambda)\right]^n
\end{multline*}
where we have used the inequality $\sum_1^k x_i^n \geq k^{1-n}
(\sum_1^k x_i)^n$ for $n \geq 1$ and $x_i \geq 0$.

By \eqref{press-conv} we have
\begin{equation*}
\frac{1}{N_{k-1}} \mathbb E\left[\left|\log
Z^{A_{N_{k}}}_{N_{k-1}}(\beta, \lambda) - \log
Z^{A_{N_{k-1}}}_{N_{k-1}}(\beta, \lambda)\right|\right] \underset{k
\rightarrow \infty}{\longrightarrow} 0
\end{equation*}
Therefore the truncated sequence of `pressures' $-\frac {1}{N_k}
\log Z^{A(\delta_{N_k}, \epsilon_{N_k})}_{N_k}(\beta, \lambda)$ is
nearly decreasing $a.s.$  It follows that the truncated
thermodynamic limits exist almost surely.

Let $X_k= -\frac{1}{N_k} \log Z_{N_k}^{A_{N_k}}(\beta, \lambda)$.
Since \{$\E \left[X_k\right]$\} is uniformly bounded, the limit is
finite a.s. Another application of the concentration inequality
\eqref{conc2} along with the Borel-Cantelli lemma implies that this
limit is a non-random constant $f(\beta, \lambda)$.  Further, it is
now a small matter to prove convergence of the truncated quenched
averages along these subsequences.  The bound \eqref{conc2} implies
that $X_k$ are uniformly integrable, i.e.
\begin{equation*}
\lim_{\kappa \rightarrow \infty} \sup_k \E \left[|X_k| \1_{|X_k|
\geq \kappa}\right] = 0,
\end{equation*}
so we may conclude $\E
\left[X_k\right] \rightarrow f(\beta, \lambda)$ as well. Finally, as
$-\epsilon_{N_k} \log \delta_{N_k} = \frac{N_k^{\ffrac 14}}{2
\beta^2}$, the arguments given above imply that $\lim_{k}
\E\left[p_{N_k}(\beta,\lambda)\right] = f(\beta, \lambda).$ Another
application of the concentration inequality then implies that
$p_{N_k}(\beta,\lambda) \rightarrow f(\beta, \lambda)\; \text{a.s.}$

It remains to prove uniqueness of this limit and then convergence
along arbitrary subsequences.  However these statements follow from
fairly standard arguments based on the above results.
\end{proof}

\section*{Acknowledgments}
\noindent The author would like to thank Marek Biskup for a number
of suggestions (in particular with respect to the universality
results), Shannon Starr for numerous useful discussions and Lincoln
Chayes for his help in preparing this manuscript.  This research was
partially supported by the NSF grants DMS-0306167, DMS-0301795 and
DMS-0505356.


\begin{thebibliography}{AA}

\bibitem{Aizenman-Nachtergaele}
M.~Aizenman and B.~Nachtergaele, \textit{Geometric aspects of
quantum spin states}, Commun. Math. Phys. \textbf{164} (1994), no.
1, 17--63.

\bibitem{A-S-S}
M.~Aizenman, R.~Sims, S.~Starr, \textit{Extended variational
principle for the Sherrington-Kirkpatrick spin-glass model.} Phys.
Rev. B   \textbf{68} (2003) 214403

\bibitem{B-H-G}
R.~Bhatt, D.~Huse, M.~Guo, \textit{Quantum Griffiths singularities
in the transverse-field Ising spin glass}, Phys. Rev. B \textbf{54}
3336-3342 (1996).

\bibitem{C-H}
Carmona, P.; Hu, Yueyun \textit{Universality in Sherrington
Kirkpatrick's spin glass model}, Annales de l'Institut Henri
Poincar\'{e} (B) Probability and Statistics, \textbf{42}, (2006),
no. 2, 215--222,  Elsevier

\bibitem{Con-Pul}
P.~Contucci, C.~Giardina, J.~Pule, \textit{Thermodynamic Limit for
Finite Dimensional Classical and Quantum Disordered Systems} Reviews
in Mathematical Physics, \textbf{16}, (2004), N. 5, 629--637

\bibitem{Cug-Grem}
L.~Cugliandolo, D.~Grempel, C.~da Silva Santos,
\textit{Imaginary-time replica formalism study of a quantum
spherical p-spin-glass model} Phys. Rev. B \textbf{64}, 014403
(2001).

\bibitem{FILS}
J.~Fröhlich, R.~Israel, E.~Lieb, B.~Simon, \textit{Phase Transitions
and Reflection Positivity. I. General Theory and Long Range Lattice
Models}, Commun. Math. Phys. \textbf{62},  (1978), 1--34

\bibitem{G-P-S}
A.~Georges, O.~Parcollet, S.~Sachdev, \textit{Mean field theory of a
quantum Heisenberg spin glass} Physical Review Letters \textbf{85},
840 (2000)

\bibitem{Guerra}
F.~Guerra, \textit{Broken replica symmetry bounds in the mean field
spin glass model}, Commun. Math. Phys. \textbf{233} (2003), no. 1,
1--12.

\bibitem{Guerra-Ton-1}
F.~Guerra, F.L.~Toninelli, \textit{The thermodynamic limit in mean
field spin glass models}, Commun. Math. Phys. \textbf{230} (2002) 1,
71-79

\bibitem{Guerra-Ton-2}
F.~Guerra,  F.L.~Toninelli, \textit{The infinite volume limit in
generalized mean field disordered models} Markov Process. Related
Fields  \textbf{9} (2003), no. 2, 195--207

\bibitem{MPV}
M.~M\'ezard,G.~Parisi, and M.A.~Virasoro, \textit{Spin glass theory
and beyond}, World Scientific Lecture Notes in Physics, vol.~9,
World Scientific Publishing Co., Inc., Teaneck, NJ, 1987.

\bibitem{N-S-1}
C.M.~Newman, D.L.~Stein,\textit{Are there incongruent ground states
in 2D Edwards-Anderson spin glasses?},  Commun. Math Phys.
\textbf{224} (2001), 205--218

\bibitem{N-S-2} C.M.~Newman, D.L.~Stein, \textit{Ordering and broken symmetry in short-ranged spin glasses},
J. Phys.: Condens. Matter \textbf{15} (2003), R1319-R1364

\bibitem{N-S-3}
C.M.~Newman, D.L.~Stein, \textit{Nonrealistic behavior of mean field
spin glasses}, Phys. Rev. Lett. \textbf{91} (2003)

\bibitem{Panch}
D.~Panchenko,  \textit{Free energy in the generalized
Sherrington-Kirkpatrick mean field model} Rev. Math. Phys.
\textbf{17} 2005 No. 7

\bibitem{Parisi}
G.~Parisi, \textit{Field theory, disorder and simulations}, World
Scientific Lecture Notes in Physics, \textbf{49}, World Scientific
Publishing Co., Inc., River Edge, NJ, 1992.

\bibitem{Rit}
F.~Ritort, \textit{Quantum critical effects in mean-field glassy
systems} Physical Review B, \textbf{55} (1997) 14096--14099

\bibitem{Sach-Ye}
S. Sachdev, J. Ye,  \textit{Gapless spin-fluid ground state in a
random quantum Heisenberg magnet}, Physical Review Letters
\textbf{70}, 3339 (1993);

\bibitem{deSanc}
L.~De Sanctis, \textit{Structural Properties of the Disordered
Spherical and Other Mean Field Spin Models}, Journal of Statistical
Physics, Jan 2006, Pages 1 - 19, DOI 10.1007/s10955-006-9167-y, URL
http://dx.doi.org/10.1007/s10955-006-9167-y

\bibitem{Talagrand-book}
M.~Talagrand, \textit{Spin Glasses: A Challenge for Mathematicians.
Cavity and Mean Field Models}, A Series of Modern Surveys in
Mathematics, vol~46. Springer-Verlag, Berlin, 2003.

\bibitem{Talagrand-paper}
M.~Talagrand, \textit{The Parisi formula}, Ann. of Math. (2)
\textbf{163} (2006), no. 1, 221--263.



\end{thebibliography}
\end{document}